\documentclass[prd,preprintnumbers,12pt]{revtex4}
\usepackage{epsfig}
\usepackage{graphicx}
\usepackage{bm}

\newcommand{\be}{\begin{equation}}

\newcommand{\ee}{\end{equation}}
\newcommand{\bea}{\begin{eqnarray}}
\newcommand{\eea}{\end{eqnarray}}

\begin{document}
\preprint{{\bf BARI-TH 494/04}}
\title{Magnetic properties of the
Larkin-Ovchinnikov-Fulde-Ferrell superconducting phase}

\author{R. Casalbuoni}\email{casalbuoni@fi.infn.it}
\affiliation{Department of Physics, University of Florence, and
INFN-Florence , Italy}
\author{R. Gatto} \affiliation{Department of
Physics, University of Geneva, Switzerland}
\author{M. Mannarelli}\email{mannarelli@ba.infn.it}
 \affiliation{Department of Physics, University of Bari, and
INFN-Bari , Italy}
\author{G. Nardulli} \email{giuseppe.nardulli@ba.infn.it}
 \affiliation{Department of Physics, University of Bari, and
INFN-Bari , Italy}
\author{M. Ruggieri}\email{marco.ruggieri@ba.infn.it}
 \affiliation{Department of Physics, University of Bari, and
INFN-Bari , Italy}
\date{\today}
\begin{abstract}
We compute, at the first order in the fine structure constant, the
parameters of the electromagnetic Lagrangian for the inhomogeneous
Larkin-Ovchinnikov-Fulde-Ferrell phase in Quantum Chromodynamics
(QCD) and in condensed matter. In particular we compute for
 QCD with two flavors the dielectric  and the magnetic
permeability tensors, and for condensed matter superconductors the
penetration depth of external magnetic fields.
\end{abstract}
\maketitle
\section{Introduction\label{sec:1}}

The aim of this paper is to compute, at the first order in the
fine structure constant, the parameters of the electromagnetic
Lagrangian for the inhomogeneous Larkin-Ovchinnikov-Fulde-Ferrel
(LOFF) \cite{LO,FF} phase in QCD and in condensed matter. In
particular we compute in 2 Flavor QCD the dielectric constant and
the magnetic permeability. For the condensed matter superconductor
we compute the penetration depth of an external magnetic  field at
$T=0$. The inhomogeneous superconductive LOFF phase was introduced
in the context of ordinary superconductors forty years ago.
 In
the original papers \cite{LO,FF} this phase was discussed for weak
ferromagnetic materials with an exchange interactions produced  by
the presence of paramagnetic impurities. It was shown that the
spin splitting generates a separation of the Fermi surfaces. This
separation, here denoted as $\delta\mu$,  is proportional in
metallic superconductors to the magnetic field $H$. For large
enough $\delta\mu$, beyond the so called Clogston-Chandrasekhar
limit $\sim\Delta_0/\sqrt 2$ ($\Delta_0$ the gap for the
homogeneous BCS case) \cite{clogston,chandrasekhar},  it can be
energetically favorable  for two electrons to form a pair with non
vanishing total momentum $|2{\bf q}|$. The main effect is an
inhomogeneous gap with a space modulation which depends on its
plane wave decomposition.
 The LOFF phase can exist also in QCD as a
particular realization of color superconductivity at non
asymptotic densities, due to difference in Fermi momenta, as
arising from different quark masses and from $\beta$ equilibrium
in dense quark matter. Translation and rotation invariance are
broken and the space dependence of the order parameter may be that
of a crystal. Such a crystalline phase of QCD might occur in
compact stars, and suggestions exist that it may explain the
variation patterns in the pulsars rotation periods (glitches)
\cite{Alford:2000sx}.

The range of densities where the LOFF phase might be energetically
favored is still matter of debate. In a recent paper
\cite{Alford:2004hz} the intermediate density region has been
studied and  the possible spatially uniform candidate phases have
been examined. The conclusion is in favor of a gapless CFL phase
(gCFL) and, for immediately lower densities, in favor of the LOFF
phase, based on the indications of the calculation in Ref.
\cite{Casalbuoni:2004wm}. These results, if confirmed, would make
more likely the occurrence of the LOFF phase at the pre-asymptotic
densities of compact stars.

An important point is the form of the condensate. Recent analyses
\cite{Casalbuoni:2004wm,Bowers:2002xr} point to cubic structures
as the energetically favored form of the condensate. They are the
body-centered cube (bcc) and the face-centered cube (fcc),
obtained summing 6 or 8 plane waves pointing to the faces or the
vertices of a cube. The bcc
 structure seems the dominant one for
$\delta\mu$ near the Clogston limit \cite{Casalbuoni:2004wm}; for
larger values of $\delta\mu$ the fcc structure is favored,
\cite{Bowers:2002xr,Casalbuoni:2004wm}.
 Therefore here we consider
three different structures, i.e. the one plane wave case
(Fulde-Ferrell phase) and the two cubic structures.
 We perform our study in a
well defined approximation, not based on the Ginzburg-Landau
approach, but valid for $\Delta$ not too small
\cite{Casalbuoni:2004wm}. This approximation is based on a
convenient average over the sites of the crystalline structure
defined by the condensate. The result can be described in terms of
a multi-valued gap function possessing $P$ branches, where $P$ is
the number of plane waves defining the crystal. Each of these
branches corresponds to a gap $k\Delta$, $k=1,\cdots,P$ with
$\Delta$ the constant gap factor appearing in the LOFF condensate.

In Sections  \ref{sec:2} and \ref{sec:3} we review the formalism
employed to describe the LOFF phase and the results obtained by
the approximation of ref. \cite{Casalbuoni:2004wm}. In  Section
\ref{sec:4} we discuss the LOFF phase in QCD. We discuss the
problem of the Meissner mass and the determination of the
parameters (dielectric constant and magnetic permeability) of the
Lagrangian for the electromagnetic field. Differently from the
homogeneous two flavor case (2SC) \cite{Rischke:2000cn}, we find a
correction not only for the dielectric constant, but also for the
magnetic permeability. Since in QCD there is a rotated $U(1)$ that
is conserved \cite{Alford:1998mk}, this implies a constraint on
our calculation scheme since the Meissner mass must vanish. We use
this result in Section \ref{sec:4} where we consider the LOFF
phase in  condensed matter. In this case  the Meissner mass does
not vanish and in general the magnetic field $H$ should be
expelled. This has been discussed in
  the  LOFF superconductor
\cite{LO} within the Ginzburg-Landau approximation for the case of
a gap with a space modulation $\sim\Delta\cos2qz$ and in \cite{FF}
for the one-plane-wave case. In this Section we consider other
crystalline structures and the region near the Clogston limit, far
away from the second order transition point.

\section{General formalism \label{sec:2}}In this Section we
 briefly review the formalism we employ to
describe inhomogeneous color superconductivity in QCD;
modifications for condensed matter applications will be discussed
in Section \ref{sec:4}.  We consider QCD with  two massless quarks
having different chemical potentials $\mu_1$ and $\mu_2$ and we
suppose that $\delta\mu=|\mu_1-\mu_2|/2$ is slightly larger than
the Clogston-Chandrasekhar \cite{clogston,chandrasekhar} limit
$\Delta_0/\sqrt 2$, where $\Delta_0$ is the value of the gap for
the homogeneous BCS phase. We work in zero temperature high quark
density limit, which means that $\mu=(\mu_1+\mu_2)/2\gg
\delta\mu$. In these hypotheses the system can be supposed to be
in the LOFF phase characterized by the following pattern of
condensation:
\begin{equation}
\left<\psi_{i\alpha}\,C\,\psi_{j\beta}\right>=\Delta\,\sum_{m=1}^{P}\,e^{2iq{\bf
 n_m}\cdot{\bf r}}\epsilon^{\alpha\beta 3}\,\epsilon_{ij}
\label{LOFFansatz}
\end{equation}
where $\alpha,\beta$ are color indices, $i,j$ are flavor indices
and $2q \bf  n_m$ is the total momentum of the Cooper pair. We
will consider below three cases. The first is the one-plane wave
Fulde Ferrel state with $P=1$ (and $\bf n$ along the $z$-axis). In
the second case we take $P=6$ with the six unit vectors $\bf n_m$
pointing to the six faces of a cube (bcc). Finally we consider the
case $P=8$ with the eight unit vectors $\bf n_m$ pointing to the
eight vertices of a cube (fcc). With the choice of phases for the
plane waves as in Eq. (\ref{LOFFansatz}) the symmetry of the
condensate both for $P=6$ and $P=8$ corresponds to the cube group.

The reason to discuss only these cases is based on the results of
\cite{Casalbuoni:2004wm}. Here it is shown that the bcc is the
energetically favored structure in the $\delta\mu$ interval
$(0.707\Delta_0-0.95\Delta_0)$, while for $\delta\mu$ in the
interval $(0.95\Delta_0-1.12\Delta_0)$ the fcc dominates. The
approximation used in \cite{Casalbuoni:2004wm} is based on the so
called High Density Effective Theory (HDET)
\cite{Hong:1998tn,Hong:1999ru,Casalbuoni:2000na,
Casalbuoni:2001ha,Nardulli:2002ma,Schafer:2003jn} and on an
averaging procedure of the original Lagrangian over a region of
the size of the lattice cell. In the HDET formulation one
decomposes the fermion momentum $p^\mu$ in its hard part
 $\mu v^\mu$ and a residual momentum $\ell^\mu$, i.e. $ p=\mu v+\ell$
 where $v^\mu=(0,{\bf v})$; ${\bf v}$ is the fermion  velocity and one neglects in $\ell$
 the transverse part  writing
 $\ell=(\ell_0,\ell_\parallel{\bf v})$,
 with $\ell_{\parallel} ={\bm\ell}\cdot{\bf v}$. Since
 $\mu_1\neq\mu_2$ we have two velocities here, but one can prove
 that in the large $\mu$ limit, ${\bf v_1}=-{\bf v_2}+{\cal
 O}(\delta\mu/\mu$).
 The momentum decomposition allows to define velocity-dependent fields,
 whose Fourier transform depends on $\ell$.
 The averaging procedure substitutes the inhomogeneous gap $\Delta({\bf r})\propto \sum\exp(2iq{\bf n_m \cdot r})$
 with a  function of $\ell$ and $\bf v$; the whole approach is justified
 if the velocity dependent fields are slowly varying over regions of the order of the lattice size.
Therefore this Lagrangian can only describe soft momenta. For more
details see \cite{Casalbuoni:2004wm}.

Let us write the fermion propagator in this approach. Since we
have four degrees of freedom (two flavors and two colors) we can
use a compact notation introducing a base of velocity-dependent
fermion fields $\psi_A$ with $A=1,\dots ,4$. In this base the
quark propagator assumes the form
\begin{equation}
S_{AB}({\bf v},\ell)=\frac{1}{D({\bf v},\ell)}\left(
\begin{array}{cc}
  \tilde{V}\cdot l\,\delta^{AB} & -\Delta_{AB} \\
  -\Delta_{AB} & V\cdot l\,\delta^{AB}
\end{array}
\right)\label{fermPROP}
\end{equation}
with $A,B=1,\dots,4$. Here $V=(1,{\bf v})$, $\tilde V=(1,-{\bf
v})$,   the gap matrix is
\begin{equation}
{\Delta}_{AB}=\Delta_{E}({\bf v},\ell_0) \left(
\begin{array}{cccc}
  0 & 0 & 0 & 1 \\
  0 & 0 & -1 & 0 \\
  0 & -1 & 0 & 0 \\
  1 & 0 & 0 & 0
\end{array}
\right) \,   \label{gapMatr2SC}
\end{equation} and \be D({\bf v},\ell)= V\cdot\ell\,\tilde V\cdot
\ell-\Delta^2_{E}({\bf v},\ell_0)\,  .\ee The effective gap is
given by  \be\Delta_{E}({\bf
v},\ell_0)=\sum_{m=1}^P\Delta_{eff}\left({\bf v\cdot\bf
n_m},\ell_0\right)\label{eq13}\, , \ee with  \be \Delta_{eff}({\bf
v\cdot n}, \epsilon)=\Delta\theta(E_u)\theta(E_d) \label{delta2}\
.\ee Here \be E_{u,d}=\pm\delta\mu\mp q{\bf v} \cdot{\bf n}
+\epsilon \label{dispersionFF}\ \ee  are the quasi-particle
dispersion laws and $\epsilon$ is the value of the energy at the
pole of the propagator.

\section{ $ \tilde U(1)$ gauge invariance and  the  LOFF phase of QCD\label{sec:3}}

The quark pair condensate (\ref{LOFFansatz}) breaks the
electromagnetic group $U(1)_{em}$ since the pairs have total
non-zero electric charge.   There exists however a group which we
call $\tilde U(1)$ generated by a linear combination of the
electromagnetic charge $Q$ and the $T_8 (=\lambda_8/2)$ color
generator
\begin{equation}
\tilde{Q}\equiv\tilde{Q}_{ij}^{\alpha\beta}=Q_{ij}\otimes
I^{\alpha\beta}-\frac{1}{\sqrt{3}}I_{ij}\otimes T_8^{\alpha\beta}
\label{QtildeDEF}
\end{equation}
that remains unbroken as far as the transverse degrees of freedom
are concerned. The residual symmetry  embodied by
Eq.(\ref{QtildeDEF}) implies a mixing angle  $\theta$ between the
photon   $\bf A$ and gluon $ \bf G^8 $:\begin{equation}
\cos\theta=\frac{g\,\sqrt{3}}{\sqrt{3\,g^2+e^2}} \label{mixing}\ ;
\end{equation}
The {\em in-medium} vector potential fields $\bf \tilde{A}$ and
$\bf \tilde{G}^8$ are then given by
\begin{eqnarray}
\tilde{A}_i&=&-\sin\theta\,G^8_i+\cos\theta\,A_i\ ,
\label{inmediumpho}\\
\tilde{G}^8_i&=&\cos\theta\,G^8_i+\sin\theta\,A_i
\label{inmedium8}\ .
\end{eqnarray}

This phenomenon is similar to what happens in the two-flavor
superconducting phase of QCD, the so-called 2SC model, where the
quark pair condensate has the same color and flavor dependence of
(\ref{LOFFansatz}), but with $q=0$.  These results indeed do not
depend on the space modulation of the condensate. Gauge invariance
under $\tilde U(1)$ implies that the polarization tensor $
\Pi_{ij}(p)$ of the $\tilde A_i$ field vanishes for zero external
momentum $p=0$. To check this result at the one loop level in the
HDET   we consider the self-energy and tadpole diagrams (see e.g.
Fig.1 in \cite{Casalbuoni:2001ha}) . They are computed by the
propagator (\ref{fermPROP}) and the effective interaction
Lagrangians
\begin{eqnarray}
{\cal L}_{qq \tilde A}&=&i\,\tilde
e\,\sum_{\bf{v}}\,\chi^\dagger_{A}\,\left(
\begin{array}{cc}
 - {\bf v\cdot}{\bf \tilde{A}}\,\tilde{Q}_{AB}
& 0 \\
  0 &  -
{\bf v\cdot}{\bf \tilde{A}}\,\tilde{Q}_{AB}
\end{array}
\right)\,\chi_{B} \label{lagra1}\ ,\\ {\cal L}_{qq \tilde A \tilde
A}&=&-\tilde e^2\sum_{\bf{v}}\,\chi^\dagger_{A}\,\left(
\begin{array}{cc}
  \frac{\tilde{Q}^2_{AB}}{2\,\mu+\tilde{V}\cdot \ell} & 0 \\
  0 & \frac{\tilde{Q}_{AB}^2}{2\,\mu+V\cdot \ell}
\end{array}
\right)\,\chi_{B}\,P^{ij}\,\tilde{A}_i\,\tilde{A}_j
\label{lagraTAD}
\end{eqnarray}  where $\tilde{e}=e\cos\theta$,
\begin{equation}
P_{\mu\nu}=g_{\mu\nu}-\frac{1}{2}\left(V_\mu\,V_\nu +
\tilde{V}_\mu\,\tilde{V}_\nu\right) \ ,
\end{equation}
and $\tilde Q=1/2\,{\rm diag}(+1,+1,-1,-1)$.

For any crystalline structure we have the result \bea
i\,\Pi_{ij}(p)
     &=&
\frac{\tilde{e}^2\,\mu^2}{12\,\pi^3}\,
\int\frac{d\bf{v}}{4\pi}\int\!d^2\ell\, v_iv_j
     \frac{V\cdot \ell V\cdot (\ell+p)+ \tilde{V}\cdot
     \ell \tilde V\cdot (\ell+p)-2\,\Delta_{E}^2}{D(\ell)D(\ell+p)}\cr
&-&\frac{\tilde{e}^2}{6 \pi^3}\,
\int\frac{d\bf{v}}{4\pi}\int\!d^2\ell\,
P_{ij}\left[(\mu+\ell_\parallel)^2 \frac{\tilde{V}\cdot
\ell}{(2\,\mu + \tilde{V}\cdot
 \ell )\,D(\ell)}\, +\,(V\rightarrow\tilde{V})\right]
\label{seTADdiagram}\ . \eea
 The former term on the r.h.s. is  the contribution of the self-energy
 graph, the latter term is from the tadpole diagram.

 For the Fulde Ferrel one plane wave case we can distinguish two
 contributions, one from the pairing region, and the other from
 the blocking region. The pairing region is defined by
 ($\xi=\ell_\parallel$):
 \be\label{PRFF} {\cal P}_1= \left\{ (\xi,{\bf v}) \,|\,\Delta= \Delta_{eff}
 ({\bf v\cdot n},\epsilon)|_{\epsilon=
\sqrt{\xi^2+\Delta^2}}\right\} \ .\ee The blocking region ${\cal
P}_0$ corresponds  to $E_u<0$ or $E_d<0$ or both.
  We have heuristically tested the vanishing of the Meissner mass
 in this case by computing numerically
 $\tilde\Pi_{ij}(0)$ for the following values of the parameters
 $\mu=400$ MeV, $\delta\mu=\delta\mu_1=\Delta_0/\sqrt 2$ and
 taking the values of
 $\delta\mu/q=0.78$ and of $\Delta=0.24 \Delta_0$ that minimize the free energy \cite{Casalbuoni:2004wm}. We find $\tilde\Pi(0)$ of the order of $10^{-3}$, while the
 two individual contributions are of the order of 1, which points
 to an almost complete cancellation.
 This result was expected on the basis of gauge invariance and
 the rather accurate approximation employed to get (\ref{dispersionFF}), i.e.
 $\delta\mu/\mu\ll 1$.

Next we consider the case of several plane waves with wave vectors
with the same modulus $q$, but  directed along the directions $\bf
n_m\,$, $m=1,\cdots,P$. Here our formalism is based on the
approach discussed in Section \ref{sec:2}. We have different
regions ${\cal P}_k$ where pairing is possible. They are defined
as follows \be {\cal P}_k=\{({\bf v},\xi)\,|\,\Delta_E({\bf
v},\epsilon)=k\Delta\}\,,\hskip1cm k=1,\cdots,P\ee  where
$\Delta_{E}$ is in (\ref{eq13}) and one uses (\ref{dispersionFF})
with ${\bf n}\to{\bf n_m}$. This approximation is the result of an
averaging procedure described in detail in
\cite{Casalbuoni:2004wm} and is valid for $\xi$ of the order of
$q$ or smaller.  Now in the self energy term in the r.h.s of
(\ref{seTADdiagram}) the relevant contribution  in the $\xi$
integration  comes from the small $\xi$ region; therefore the
approximation is adequate. On the other hand in the second term
(tadpole contribution) the hard modes dominate and the
approximation is no longer valid. Therefore we use $\tilde U(1)$
gauge invariance to get information on the main features of the
pairing regions for large $\xi$. Notice that the sum over $k$
arises because  plugging (\ref{eq13}) into (\ref{seTADdiagram})
one has  several terms, corresponding to different values assumed
by the gap: $k\Delta$ ($k=1,\cdots,P$). Imposing gauge invariance
one gets \bea 0&=&\sum_k\left\{\frac 1 2\int_{{\cal P}_k}
\frac{d\xi\,d\bf{v}}{4\pi}\frac{k^2\Delta^2}{(\xi^2+k^2\Delta^2)^{3/2}}\,
\right.\cr &+&\left.\int_{\tilde{\cal
P}_k}\frac{d\xi\,d\bf{v}}{4\pi\sqrt{\xi^2+k^2\Delta^2}} \left(
\frac{(1-\xi/\mu)^2(\sqrt{\xi^2+k^2\Delta^2}+\xi)}{
\sqrt{\xi^2+k^2\Delta^2}+\xi-2\mu}\,+(\xi\leftrightarrow-\xi)
\right) \right\} \label{20} \eea where $\tilde{\cal P}_k$ is the
region of the phase space where $\Delta_E=k\Delta$, but $\xi\sim
\mu$. Even if this equation does not determine $\tilde{\cal P}_k$,
it provides sufficient information  to compute the Meissner mass
in ordinary LOFF superconductors. An application of this result is
discussed in the next Section.

For small external momentum we get from Eq.(\ref{seTADdiagram})
and from the condition of vanishing Meissner mass, that
\begin{equation}
i\Pi(p)_{ij}\approx-\frac{\tilde
e^2\mu^2}{12\pi^3}\left[A_{ij}p_0^2 + B_{ij}^{kl}p_k p_l\right]
\label{smallp1}
\end{equation}
where we have defined \be A_{ij}=\sum_{k=1}^P \int_{{\cal
P}_k}\!\frac{d^2\ell d{\bf v}}{4\pi}v_i v_j \frac{2}{D^2(l)}
\,,\hskip.5cm B_{ij}^{kl}=\sum_{k=1}^P \int_{{\cal
P}_k}\!\frac{d^2\ell d{\bf v}}{4\pi}v_i v_j v_k
v_l\left[\frac{2}{D^2(l)} + \frac{4\Delta_E^2}{D^3(l)} \right] \,
. \label{23} \ee
 For the FF state we have two
independent transverse tensors, $\Pi^T_1=\Pi_{11}$ and
$\Pi^T_3=\Pi_{33}$, while for both cubic structures we have
\be\Pi^T(p)=\frac{1}{2}\left(\delta_{ij}-\frac{p_i p_j}{{\bf
p}^2}\right)\Pi_{ij}(p)\label{eq:24}\ee as in the homogeneous
case. The Lagrangian for the rotated photon can be written in the
form
\begin{equation}
{\cal L}=\frac{1}{2}\left(\epsilon_{ij}E_i E_j -
\frac{1}{\lambda_{ij}}B_i B_j\right),
\end{equation} where
$ \epsilon_{ij}$ and $\lambda_{ij}$ can be obtained from Equations
(\ref{23}).
 In general we have
\be \epsilon_{ij}= \left(1+f_j(\delta\mu,\Delta_0)\frac{\tilde
e^2\mu^2}{18\pi^2\Delta^2}\right)\,\delta_{ij}\ ,\hskip.3cm
\lambda_{ij}^{-1}= \left(1+g_j(\delta\mu,\Delta_0)\frac{\tilde
e^2\mu^2}{18\pi^2\Delta^2}\right)\,\delta_{ij}\ . \ee The
coefficients $f_j,g_j$ assume different values according to the
crystalline structure. At $\delta\mu=\delta\mu_1$ and $\Delta_0=
40$ MeV we have, for the one-plane-wave (FF)  $f_1=f_2 = +0.12$,
$f_3=+0.23$, $g_1=g_2 = +0.31 \times 10^{-2} $ and $g_3= +0.13
\times 10^{-3} $. For the body-centered-cube $f_{bcc}= +0.49  $
and $g_{bcc}= -0.09$. For the face-centered-cube at
$\delta\mu=0.95 \Delta_0$ we have $f_{fcc} =  +0.46 $ and
$g_{fcc}= -0.09$. For the 2SC case we have, in agreement with
\cite{Rischke:2000cn,Litim:2001mv}, $f_{2SC} =1$ and $g_{2SC} =
0$, showing  absence of corrections for the magnetic permeability
in the homogeneous phase.

\section{penetration depth in condensed matter
\label{sec:4}}As an application of the result (\ref{20}), in this
Section we give an estimate of the penetration depth of a weak
static magnetic field in an ordinary condensed-matter  LOFF
superconductor at $T=0$. We assume that the field is small enough
to produce an exchange field and a sizeable paramagnetic effect so
that the Fermi surfaces of the two pairing electrons are separated
and the  LOFF phase is formed (in particular the
Clogston-Chandrasekhar limit is reached). At the same time we
assume that the effect of the external field can be neglected in
the gap equation. This is a strong assumption, as we know that the
effect of an external magnetic field is a modulation of the gap,
see e.g. \cite{Gruenberg:1966ab}. As a matter of fact the
paramagnetic effect, which is needed to produce the separation of
the Fermi surfaces, and the diamagnetic effect, which is
detrimental to superconductivity, are in general related.
Therefore the original proposal of \cite{LO,FF} is now considered
only as an ideal case. The actual experimental activity points to
layered superconductors where one can minimize diamagnetic effects
by choosing the external magnetic field parallel to the layer. We
refer the interested reader to the specialized literature (see
e.g. \cite{Casalbuoni:2003wh} and references therein) for a
discussion. For the time being we study the effect of the magnetic
field in the idealized case where its effects on the gap equations
can be neglected.

 Let us assume that a plane surface ($yz$ plane)
divides the space into two parts, one containing the
superconductor in the LOFF phase (half-space with $x>0$) and the
other containing matter in the normal phase. At the interface the
magnetic field $\bf H$ is parallel to the $yz$ plane. We take $\bf
H$ along the $z-$axis and the vector potential $\bf A$ directed
along the $y-$axis; $\bf A$ depends only on $x$ and we assume
$\nabla\cdot{\bf A}=0$.

In condensed matter the LOFF condensate has the form similar to
(\ref{LOFFansatz}):
\begin{equation}
\langle\psi\,C\,\psi\rangle = \Delta({\bf
r})=\Delta\,\sum_{m=1}^{P}\,e^{2\,i\,q\bf{n}_m\cdot{r}}\label{LOFFcondensate}
\end{equation}
where the $\psi$ are non relativistic, two components spinor
fields describing electrons. The Lagrangian can be written as
follows \cite{Casalbuoni:2003wh}
\begin{equation}
{\cal L}=\sum_{\bf{v}}\chi^\dagger_a\, \left(\begin{array}{cc}
(V\cdot\ell+ e\,{\bf v\cdot A})\,\delta^{ab} +
\delta\mu\,\sigma_3^{ab}& -\Delta_{E}({\bf v},\ell_0)\,\delta^{ab}\\
-\Delta_{E}({\bf v},\ell_0)\,\delta^{ab}&({\tilde V}\cdot\ell-
e\,{\bf v\cdot A})\,\delta^{ab} + \delta\mu\,\sigma_3^{ab}
\end{array}\right) \,\chi_b \label{gaplagrELECTR} \, ,
\end{equation}where we have used the effective $\Delta_{E}({\bf v},\ell_0)$
approximation; $\delta\mu$ is proportional to the exchange field
acting on the electron spin and the  term $\delta\mu \sigma_3$
describes a paramagnetic coupling.
 In (\ref{gaplagrELECTR}) $a,b$
are spin indices and $\chi_a$ are Nambu-Gorkov fields. The
Lagrangian includes the coupling to the external vector potential
field ${\bf A}$.  The penetration depth is defined by
\cite{Landau2} \be \delta=\frac 1{H_0}\int^{\infty}_0 dx\, H(x)\
\, ,\ee where $H_0$ is the value of the magnetic field outside the
superconductor.  If rotational symmetry holds (the BCS and the
cubic structures) one gets, using previous hypotheses
\cite{Landau2}:
\begin{equation}
\delta\,=\,\frac{2}{\pi}\int_{0}^{\infty}\!dp\,\frac{1}{p^2-\Pi_T(p)}
\label{penDepth} \end{equation} where $ \Pi_T({{p}})$ is computed
by (\ref{eq:24}) in the static $p_0=0$ approximation; for
$\Pi_{ij}({\bf p})$ we have\begin{equation} \Pi_{ij}({\bf
p})=-\frac{e^2\,p_F\,m\,v^2}{4\,\pi^2}\sum_{k}
\,\int_{-\infty}^{+\infty}\!d\xi\!\int_{{\cal P}_k
}\!\frac{d\,{\bf{\hat v}}}{4\,\pi}\,\hat v_i\,\hat v_j\,J
\left(\xi,\beta,\,k\,\Delta\right)\,+\,\delta\Pi_{ij}
\label{generalPIcm}\ \, ,
\end{equation}
where  we have defined
\begin{equation}
J(\xi,\beta,k\,\Delta)=\frac{1}{\xi\,\beta}
\left[\frac{\xi^2-\beta\,\xi+k^2\,\Delta^2}{\sqrt{(\xi-\beta)^2+k^2\,\Delta^2}}
-
\frac{\xi^2+\beta\,\xi+k^2\,\Delta^2}{\sqrt{(\xi+\beta)^2+k^2\,\Delta^2}}\right]
\ ,\end{equation} with $\beta=\bf{p\cdot \hat v}/2$ and $\bf{\hat
v}$ is the direction of  Fermi velocity $\bf v$. The sum over $k$
in (\ref{generalPIcm}) goes from $k=0$ to $k=P$. For the BCS
homogeneous case there is one term: $k=1$. For the one-plane wave
case there are two terms, one with $k=0$ corresponding to the
blocking region, where $\Delta=0$, and the other one with $k=1$,
corresponding to the pairing region. For all the other cases the
sum  runs from $k=1$ to $k=P$. As a matter of fact, within our
approximation \cite{Casalbuoni:2004wm}, for the structures with
more than one plane wave, pairing is possible in the whole phase
space. In these cases, as discussed in Section IV.A of
\cite{Casalbuoni:2004wm}, one can identify the blocking region
with the domain where only the branch with gap $P\Delta$ of the
dispersion law contributes. The two terms correspond to the two
terms on the r.h.s. of Eq. (\ref{seTADdiagram}).

We are interested in Type II superconductors where the relevant
momenta in Eq. (\ref{penDepth}) are $p\simeq 0$. For the BCS case
one gets\be
\delta\,\approx\,\frac{2}{\pi}\int_{0}^{\infty}\!dp\,\frac{1}{p^2+m_M^2}\,=\,\frac{1}{m_M}
\label{London}\ee where $ m_M^2=\frac{e^2\,p_F^2}{3\,\pi^2}\,v $
is the squared Meissner mass in the BCS phase.

Let us consider now the LOFF phase. For the one-plane-wave case we
have
\begin{equation}
\Pi_{ij}({\bf p})=\,-m_M^2\left(\delta_{ij}-3\int_{{\cal
P}_0}\!\frac{d\,{{\bf \hat v}}}{4\pi}\hat v_i\,\hat v_j
+\,\frac{3}{4}\,\int\!d\xi\!\int_{{\cal P}_1}\!\frac{d\,{{\bf \hat
v}}}{4\,\pi}\,\hat v_i\,\hat v_j\,{J }\left(\xi,\frac{\bf{p\cdot
\hat v}}2,\Delta\right) \right)\, \, .
 \label{Pione}
\end{equation}
 The
dependence on the total momentum of the Cooper pair $2\bf q$ is in
the definition of ${\cal P}_1$. It is convenient to consider the
tensor $\tilde \Pi$ with $\bf{q}$ along the $x$-axis. The relation
between the two tensors is (sum over $k,l$)
\begin{equation}
\Pi_{ij}({\bf p})=
R_{ik}(\theta)\,R_{jl}(\theta)\,\tilde{\Pi}_{kl}({\bf p}) \, ,
\label{POLoneWAVE}
\end{equation}
where $R_{ij}(\theta)$ is the rotation matrix which brings
$\bf{q}$ along the $x$ axis, that is along $\bf p$ (the direction
of the gradient of the magnetic field).
 $\tilde \Pi$ has two independent components $
\tilde\Pi_{11}=\tilde \Pi_{22}$ and $\tilde \Pi_{33}$ and the
superconductor is characterized by two independent  penetration
lengths which we compute in the London limit. We do that for the
following values of the parameters \cite{Casalbuoni:2004wm}:
$\delta\mu=\delta\mu_1$, $z_q=0.78$ and $\Delta=0.24\,\Delta_0$.
In this case we get: $ \delta_1 \simeq 2.6 \,\delta_L$ and
 $ \delta_3 \simeq 1.4\, \delta_L$ where $\delta_L$ is the London
 penetration depth in the BCS case. Within our approximation,
 consisting in neglecting terms of the order $\delta\mu/\mu$ our
 results are compatible with those of \cite{FF}; for example for
$ \delta_3/\delta_L$ we find agreement with \cite{FF} within 10\%.

We notice that for $\Delta\to 0$, near the second order phase
transition, the pairing region vanishes, whereas the blocking
region ${\cal P}_0$ is the whole Fermi surface. From (\ref{Pione})
we see that $\Pi_{ij}(0)$ vanishes, and from (\ref{penDepth}) we
get that both $\delta_i$ diverge, which means that the FF is no
longer a superconductor, in agreement with the result of
\cite{LO}.

We can repeat the analysis for cubic crystalline structures (bcc
and fcc). In these cases one can exploit the residual discrete
symmetry and  only one penetration length is present.

 The transverse
component of the polarization tensor is obtained by
(\ref{generalPIcm}) using the appropriate set of plane waves. In
the London limit only $p\sim 0$ are relevant and one gets:
 \bea
\Pi_T (0)&=& 2 m^2_M \sum_{k=1}^P\int_{\tilde{\cal
P}_k}\frac{d\xi\,d\bf{\hat v}}{4\pi\sqrt{\xi^2+k^2\Delta^2}}
\left( \frac{(1-\xi/\mu)^2(\sqrt{\xi^2+k^2\Delta^2}+\xi)}{
\sqrt{\xi^2+k^2\Delta^2}+\xi-2\mu}\,+(\xi\leftrightarrow-\xi)
\right) \cr&=& -m^2_M \sum_{k=1}^P\int_{{\cal P}_k}
\frac{d\xi\,d\bf{\hat
v}}{4\pi}\frac{k^2\Delta^2}{(\xi^2+k^2\Delta^2)^{3/2}}\, . \eea In
the second line we have used the result expressed by Eq.
(\ref{20}). Numerically we get for the London penetration length
in the bcc case $\delta\approx 0.69\delta_L$ at
$\delta\mu=\delta\mu_1$; for the fcc case we get  $\delta\approx
0.52\delta_L$ at $\delta\mu=0.95 \Delta_0$, where, according to
\cite{Casalbuoni:2004wm} there is a transition from the bcc to the
fcc LOFF phase. \section{Conclusions} We have used the high
density effective theory formalism to  compute the low energy
properties of the electromagnetic Lagrangian of the LOFF phase in
QCD and condensed matter. We have shown that  in  QCD the rotated
photon associated to the unbroken $\tilde U(1)$ group is screened
both electrically and magnetically. We have  computed near the
Chandrasekhar-Clogston point  the dielectric tensor and the
magnetic permeability tensor for the one-plane-wave, the
body-centered cube and the face-centered cube crystalline
structures. In condensed matter we have computed  the penetration
depth of an external magnetic field. In the London limit the
penetration depth is proportional to the London penetration depth
of the BCS case with coefficients that assume different values
according to the crystalline structure.

\end{document}